\documentclass[runningheads]{llncs}
\usepackage[utf8]{inputenc}
\usepackage{graphicx}
\usepackage{xcolor}
\usepackage{multirow}
\usepackage{url} 
\usepackage{listings}
\usepackage{geometry}
\usepackage{subcaption}
\lccode`0`0
\lccode`1`1
\lccode`2`2
\lccode`3`3
\lccode`4`4
\lccode`5`5
\lccode`6`6
\lccode`7`7
\lccode`8`8
\lccode`9`9

\definecolor{mygreen}{rgb}{0,0.6,0}
\definecolor{mygray}{rgb}{0.5,0.5,0.5}
\definecolor{mymauve}{rgb}{0.58,0,0.82}

\AtBeginDocument{%
  \providecommand\BibTeX{{%
    \normalfont B\kern-0.5em{\scshape i\kern-0.25em b}\kern-0.8em\TeX}}}
    
\usepackage[acronym,toc,shortcuts]{glossaries}

\newacronym{stl}{STL}{Standar Template Library}
\newacronym{phast}{PHAST}{Parallel Heterogeneous-Architecture STL-like Template}
\newacronym{gpgpu}{GPGPU}{General Purpose Graphics Processing Unit}
\newacronym{fpga}{FPGA}{Field Programmable Gate Array}
\newacronym{asic}{ASIC}{Application-Specific Integrated Circuit}
\newacronym{dsa}{DSA}{Domain Specific Architecture}
\newacronym{dsl}{DSL}{Domain Specific Language}
\newacronym{dnn}{DNN}{Deep Neural Network}
\newacronym{hpc}{HPC}{High Performance Computing}
\newacronym{bsc}{BSC}{Barcelona Supercomputing Center}
\newacronym{api}{API}{Application Programming Interface}
\newacronym{icd}{ICD}{Installable Client Driver}
\newacronym{gemm}{GeMM}{General Matrix Multiplication}
\newacronym{pgas}{PGAS}{Partitioned Global Address Space}
\newacronym{simt}{SIMT}{Single Instruction Multiple Threads}
\newacronym{simd}{SIMD}{Single Instruction Multiple Data}
\newacronym{ann}{ANN}{Artificial Neural Network}
\newacronym{ml}{ML}{Machine Learning}

\begin{document}

\title{Using PHAST to port Caffe library: \\ First experiences and lessons learned}

\author{%
     Eduardo Jos\'e G\'omez-Hern\'andez\inst{1} \and
     Pablo Antonio Mart\'inez-S\'anchez\inst{1} \and
     Biagio Peccerillo\inst{2} \and
     Sandro Bartolini\inst{2} \and
     Jos\'e Manuel Garc\'ia\inst{1} \and
     Gregorio Bernab\'e\inst{1} 
}
\authorrunning{G. Eduardo Jos\'e et al.}

\institute{Computer Engineering Department, University of Murcia, Spain
\email{\{eduardojose.gomez,pabloantonio.martinezs,jmgarcia,gbernabe\}@um.es}\\
\and
Department of Information Engineering and Mathematical Sciences, University of Siena, Italy\\
\email{\{peccerillo,bartolini\}@dii.unisi.it}}

\maketitle

\begin{abstract}
Performance has always been a hot topic in computing. However, the viable ways to achieve it have taken many forms in the different moments of computing history. Today, technological 
limits have pushed the adoption of increasingly parallel multi-core and many-core architectures and even the use of highly specific hardware (aka Domain-Specific Architectures, or DSAs) to solve very specific problems. 

In this new context, one major problem is how to develop software once, and be able to run it on multiple accelerator architectures, seamlessly. Ideally aiming at a single programming model that can automatically target the code to different kinds of parallel architectures, allowing specific tuning with minimal, if any, changes to the source-code in order to seek performance portability. A comprehensive solution to this is still lacking.

In this work, we present the use of the PHAST Library, which allows users to code once, at a high level of abstraction and thus with high productivity, and automatically targeting different parallel devices by changing the compilation process. As a case study, we have worked on the porting of the well-known deep-learning Caffe framework. The framework has been split into different parts and some of them have been ported, obtaining a working straightforward implementation that can be run on both CPUs and GPUs.

We conclude discussing the lessons learned during the porting process, and analyzing the obtained performance in the perspective of completing the porting and expanding it to future consequent works.

\end{abstract}

\section{Introduction}
In computing, faster means better. Since the first day of the history of computing, performance has always been the main driving factor for innovation. CPUs evolved from a simple concept to huge computation monsters, as their transistor number was growing with a rate described by Moore's law. When it seemed that this trend was going to end because of the approaching of physical limits, it seemed that the performance growth itself was going to end too. However, it resisted and is still valid today thanks to a paradigm shift: the adoption of parallel architectures. The idea seemed quite simple at the same time as results were promising. The parallelism era arrived.

Nowadays, the increase of computing power comes from specialization. Instead of big powerful and power-hungry processing units, the trend is looking at \emph{specific} hardware to do \emph{specific} tasks. This specific hardware, generally called \emph{accelerator}, is able to perform computations much faster, but also using less energy. Wise linking between specific tasks and specific hardware may be difficult, but leads to otherwise unattainable results. Some researchers even see this as a new golden age for computer architecture \cite{Hennessy:2019:NGA:3310134.3282307}. However, the big problem around this new era of accelerators is to adapt the software to the specific hardware. The creation of a programming model able to take advantage of all this diversity is extremely important. This new approach is called performance portability~\cite{neely2016doe}, and it is still an open field of investigation with no definitive solution available.

A growing number of approaches have been recently proposed or improved. Big companies, like Intel (which is now launching Intel One API~\cite{intelOneApi}) are also showing interest in this topic. In this paper, we choose PHAST Library as our case study. PHAST code can be written once and targeted to different devices via a single macro. Its inner layers are implemented in CUDA C++ or \texttt{std::threads}, thus allowing targeting both NVIDIA GPUs and multi-core CPUs. 

On the application side, deep learning is of an obvious interest nowadays. Not only due to the great number of possibilities it offers, but also because of the promising future it is facing. In this work, we present a straightforward course on how to port Caffe (a deep learning framework) library using PHAST Library. These first experiences can be utilized as knowledge base for other, more complex, deep-learning cases (e.g. bigger and more complex networks).

The rest of the paper is organized as follows: Section 2 presents a background of state-of-the-art in the core and related topics of this paper, in which PHAST and Caffe can be framed. Our work is explained in Section 3, followed by the results obtained, in Section 4. Finally, Section 5 concludes the paper giving some hints for future work.

\section{Background}
\subsection{Accelerators}

After reaching the power density limit in traditional processors, a slowdown in performance improvements seemed to be unavoidable. However, a new paradigm, based on the increase of throughput through the parallel execution of multiple independent workloads, prevented said slowdown: multi-core CPUs. They quickly spread in industrial and consumer markets, growing up to the point of being responsible for the most part of the computing power in data centers.

Meanwhile, some researchers, especially in computer vision, noticed that the GPUs, that they were already using, were able to solve their problems faster. Therefore, they started using programming languages meant for graphics, like OpenGL, to implement general purpose algorithms~\cite{Bohn98kohonenfeature}.

Nowadays, GPUs are used like CPUs (GPGPU), and specific languages to use them (such as CUDA, or OpenCL) have been developed. They are able to generate graphics and compute data at high speed exploiting their \gls{simt} execution model, based on \gls{simd}, but with multi-threading. 

There are many kinds of accelerators, with GPUs being the most common. Other kinds include FPGAs and ASICs, and most of them are problem-specific. Each one of these accelerators uses its own programming language, a \gls{dsl}. Using its own \gls{dsl} for each device allows to benefit from the specific features of that device~\cite{Hennessy:2019:NGA:3310134.3282307}.

The main problem with the diffusion of accelerators to solve various problems is that each \gls{dsl} is quite different from the others. Hence, making a portable software between accelerators is impossible without having multiple source files (at least one file per target device). Having multiple versions of the same code increases the code complexity and hampers the overall code productivity and maintainability~\cite{8672426}.

\subsection{ML \& HPC}
Theoretical and mathematical models of the artificial intelligence techniques were developed in the twentieth century, but the lack of computing power prevented it from progressing. Nowadays, the performance of current hardware has become a key enabling factor for its revamping and widespread commercial adoption in various fields. For example, medical image analysis has started to implement deep learning for screening and localization of malignant zones. Additionally, other medical areas are working with these kind of techniques as well, like the analysis of the genetic information inside DNA and RNA series~\cite{Angermueller878}. The common objective is not to replace physicians with deep learning techniques, but to support them to make better diagnoses.

\gls{ml} techniques could be difficult to code and debug, therefore many frameworks have been developed to ease their use. Most of them are open-source and provide software solutions for most of the types of neural networks. The most known ones are Caffe, Caffe2, Tensorflow, Theano, PyTorch, Mxnet, and CNTK among others~\cite{8351898}.

In \gls{ml}, the training phase is very time-consuming, since it is an evaluation and optimization problem with hundreds, thousands, even millions of parameters. Therefore, the reduction of the training phase execution time is a desirable feature that frameworks should provide. Most of them are able to use NVIDIA's cuDNN, BLAS-like libraries, MKL-libraries, OpenCL, even specific libraries for custom integrated circuits, to speed up this computation. Moreover, many of them also allow other kinds of parallelism for multiple nodes, mainly using the MPI library for inter-process communication.

\subsection{PHAST}
\Gls{phast} Library~\cite{phast1,phast3,phast4,phast2} is a modern C++ programming template library based on the classic \gls{stl}-like and multi-dimensional containers, developed with to seek productivity and performance portability. It currently supports multi-core CPUs and NVIDIA GPUs, allowing the users to write expressive and concise sequential-like code that can be automatically parallelized. Its main goal is to let the programmers code using high-level programming approaches without preventing them from applying low-level optimizations, if needed, keeping the main code at a higher level of abstraction.

Similar to \gls{stl} containers, \gls{phast} provides a vector container, but also adds a matrix, cube, and grid, all of them working with a very similar interface. It also takes advantage of vector-like primitives mapping to SSE or AVX vectors.

These containers can be modified using functors, an idea borrowed by \gls{stl}. They are structs that inherit from a base functor struct and define at least operator `$()$' between their methods. There are a lot of predefined algorithms and functors, but this allows the creation of functionalities that are not defined by default in the library, without losing performance or portability.

In its roadmap, there is planned support for multiple devices in the same executable, lambda syntax, \gls{fpga} support, OpenCL backend, multi-GPU, and many other interesting features.

After writing the sequential code in the \gls{phast} way, it is possible to change the device target changing a macro and the compiler. Therefore, having two different makefiles makes the trick. The most important thing is that the code has not changed, only the compilation process.

\subsection{Caffe}
Caffe is the first \gls{dnn} framework, developed by Berkeley AI Research. Nowadays, in the production environment, Caffe is replaced by Caffe2 and pyTorch, the Caffe successors, but in other cases, TensorFlow and MXNet are the selected ones. Caffe continues to be used in research, due to the fact that it is very easy to modify, extend, or use, all of this without losing flexibility to run most of the state-of-the-art \gls{dnn} models.

This framework runs on CPU or GPU, just by changing a flag. However, this is done by having two implementations in two source-code files, one for CPU (.cpp) and on for GPU (.cu). Therefore, developers are forced to maintain two different versions of the same functionality. In this case, this is very well done and there are not a large number of differences between the files. So, it constitutes an interesting and challenging case for a single-source approach.

In the internal structure of Caffe, we have found that it is built from multiple modules that work by themselves. It is possible to classify the blocks in two parts: we call them \emph{containers} and \emph{executors}. Containers store data to be used by executors. Executors use the containers to exchange data and process it. For example, a layer gets a set of blobs, and with its own blobs, it computes the output ones.

\begin{figure}[h]
  \centering
  \includegraphics[width=0.7\textwidth]{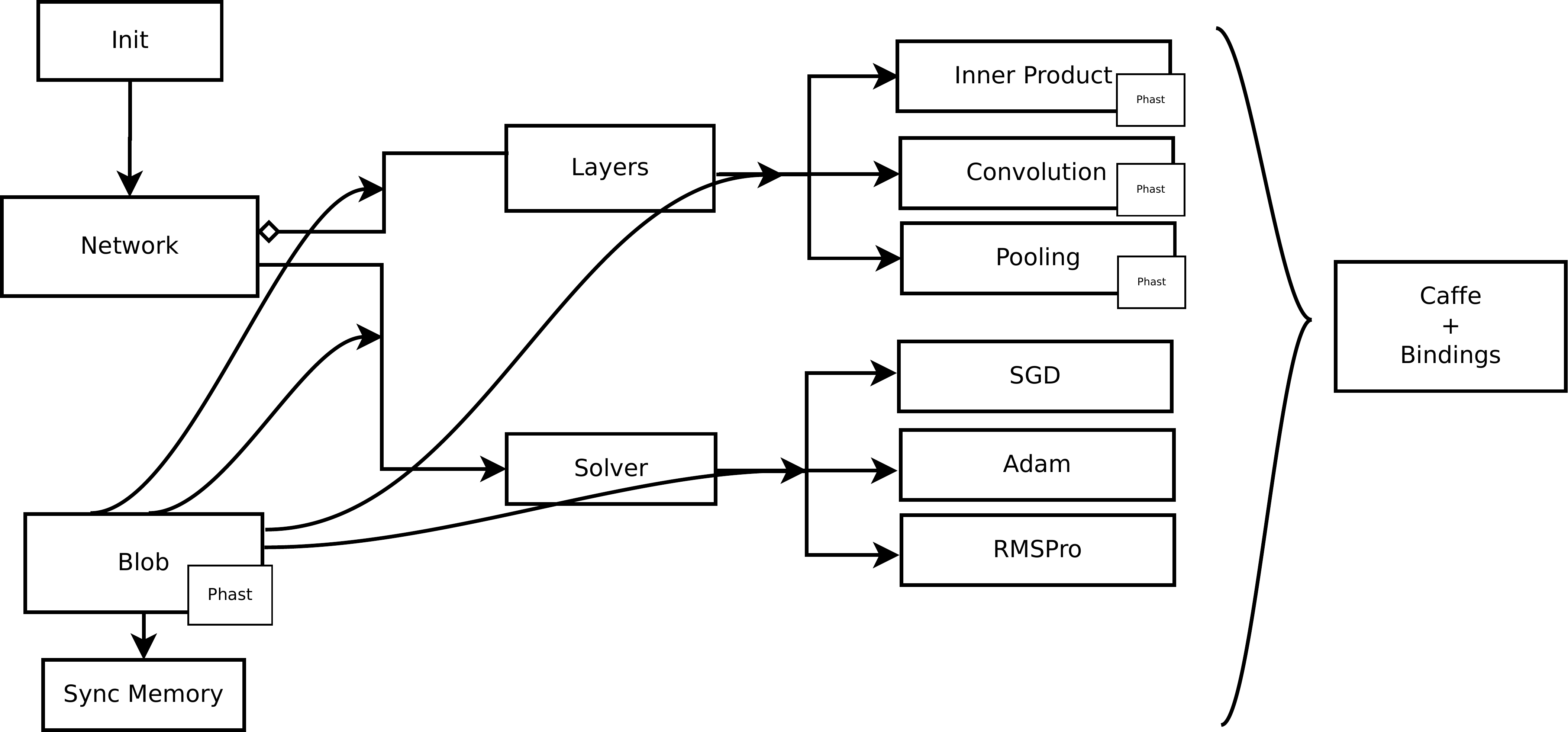}
  \caption[Caffe framework as a block diagram]{\label{fig:capha} Part of the Caffe framework as a block diagram showing communication between blocks.}
\end{figure}

A neural network has two different phases: inference and training. In the inference phase, data is passed through all the layers of the neural network in feed-forward mode to reach the last layer and get a result. But, in training mode, data is passed through all layers in feed-forward, like inference, but when it reaches the last layer of the network, that data is brought to a solver, it recalculates some values and starts the back-propagation through each layer, but in reverse order.

\section{Porting process}
In this paper, we present our port of Caffe to PHAST library, in order to be able to have one single code that can be run on both CPUs and GPUs. Our goal is to port all the blocks needed to be able to run Caffe for two LeNet variations of a sample network and MNIST and CIFAR-10 databases. The blocks we have identified for the porting are the following:

\begin{itemize}
\item \textbf{Blob}: A storage block which stores two vectors (\texttt{data} \& \texttt{diff}) used in most of the computations.
\item \textbf{InnerProduct}: One of the first layers developed in neural networks, also known as the perceptron layer. As its name suggests, it executes an inner product.
\item \textbf{Convolution}: The most common neural network layer nowadays. It is a sliding window that applies a set of filters to the input.
\item \textbf{Pooling}: Another sliding window layer that applies a simple mathematical function to reduce the size of the input.
\item \textbf{ReLU}: A simple layer that applies the ReLU mathematical function to each element. In Caffe, the leaky-ReLU version is implemented instead of a normal ReLU.
\item \textbf{Accuracy}: This is not a real layer (it is implicitly included), but it calculates the accuracy of the network for a specific set of inputs.
\item \textbf{SoftMax}: When working with classification networks, it is expected to get probabilities as an output. The SoftMax layer maps any set of numbers to probabilities that will add up to 1.
\item \textbf{SoftMax with Loss}: It is the same as the SoftMax layer, but it also computes a \texttt{loss} that can be used to know how the neural network is performing.
\end{itemize}

The more critical blocks are detailed later. We expect to run these neural networks through the ported Caffe binary in train and test mode using \gls{phast}. Therefore, we chose to modify only the blob and the lowest part of the main executors (the little "Phast" boxes in Figure \ref{fig:capha}). Furthermore, we removed the GPU code from Caffe to use the CPU version as our base for this project. In this way, we focused on the algorithm itself and not in its GPU implementation. 

\subsection{Convolution} 

\subsubsection{Feed-Forward}
\begin{figure}[ht!]
  \centering
  \includegraphics[width=0.4\textwidth]{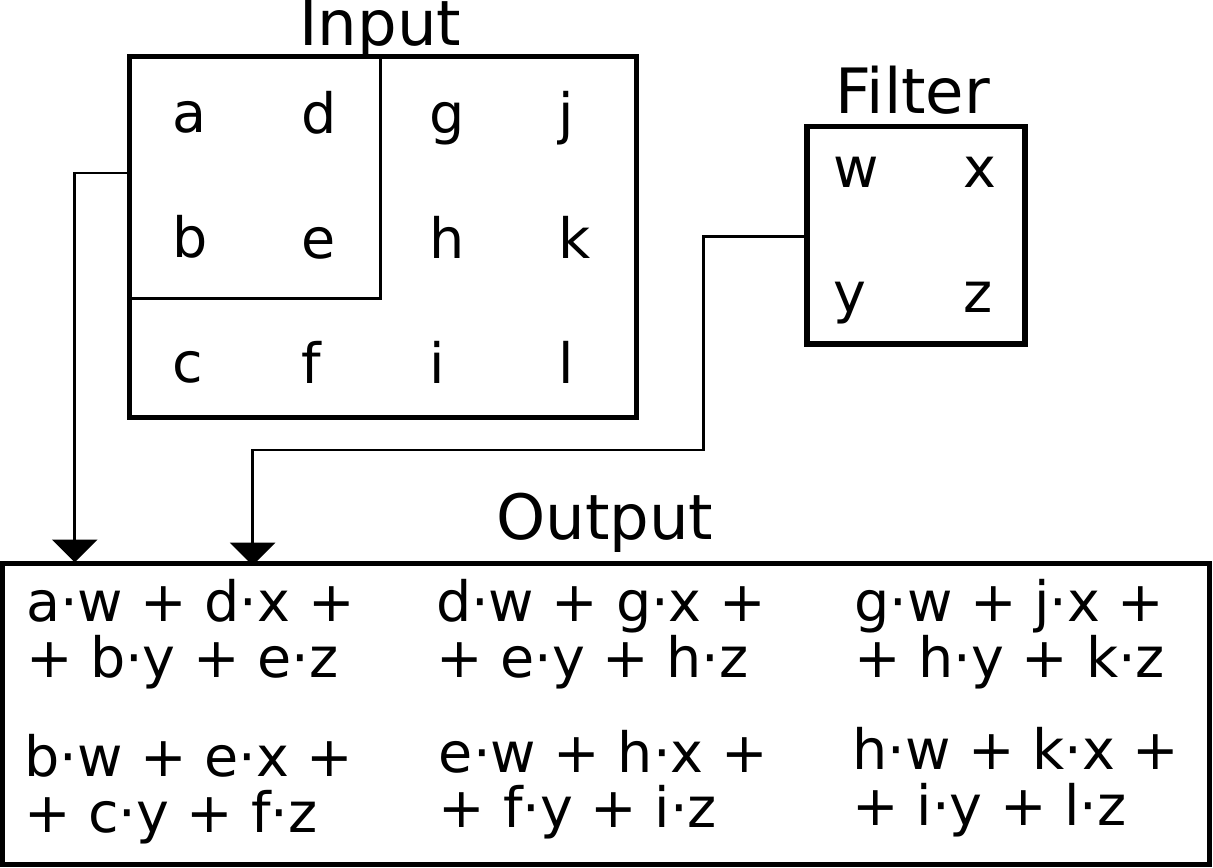}
  \caption[Classic 2-D Convolution]{\label{fig:conv1} A convolution example using a 2x2 filter with stride 1 and padding 0 over a 4x3 input matrix.}
\end{figure}

The Convolution block, or Convolution Layer, applies to the input a set of filters using a sliding window over the input. There are many ways to implement this sliding window, but the application of the filter involves the calculation of a vector inner product for each sliding window. The most common variant of Convolution is the 2-D Convolution (Figure \ref{fig:conv1}), which is a simplification of the N-D Convolution.

As our example network (LeNet) only uses 2-D Convolution, we only ported that specific variation. There is not much difference between a 2-D Convolution and a  N-D Convolution, because we use the \texttt{im2col + gemm} implementation.

\begin{figure}[ht!]
  \centering
  \includegraphics[width=0.5\textwidth]{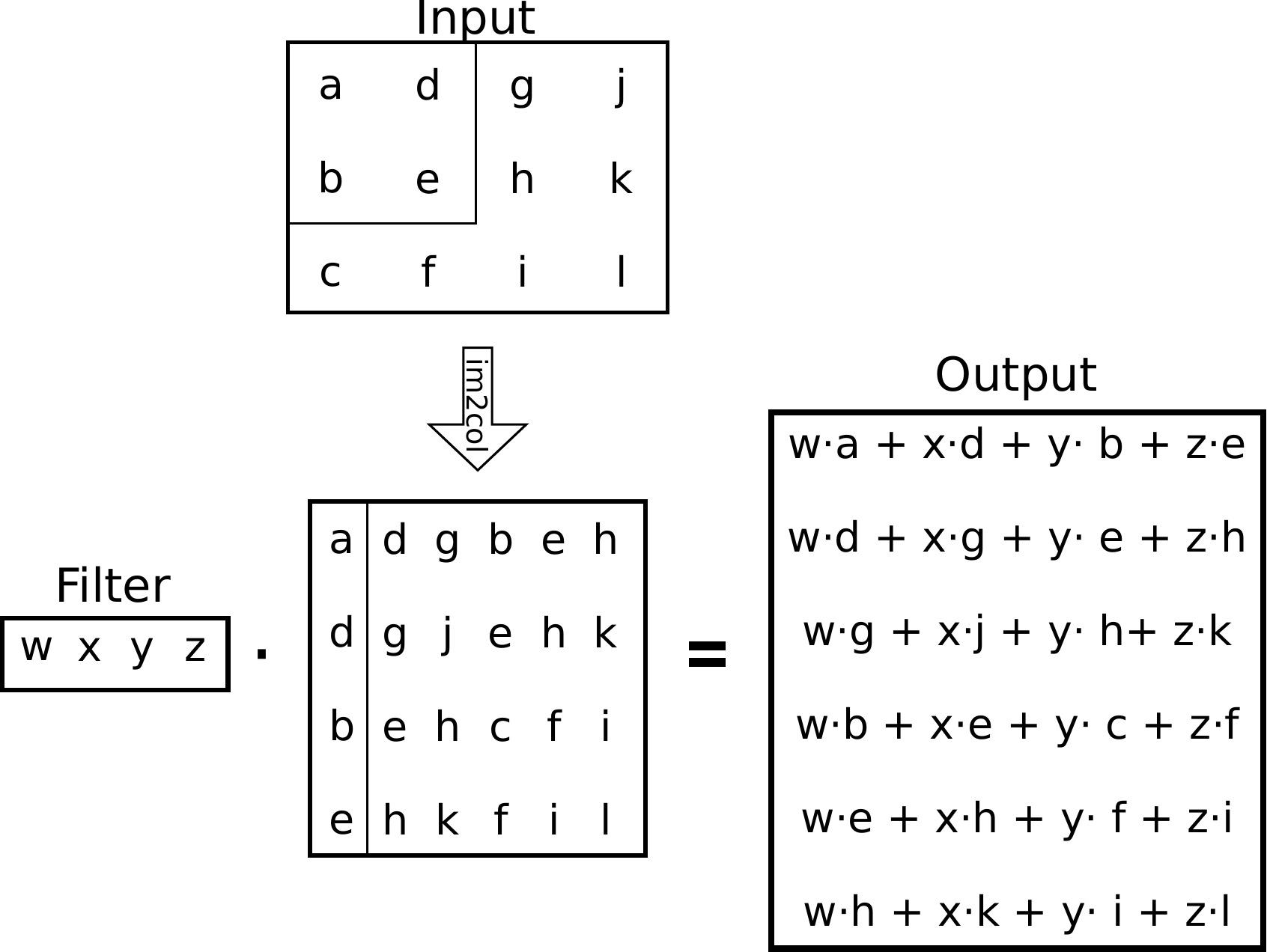}
  \caption[2-D Convolution as a GEMM]{\label{fig:conv2} Convolution as a GeMM using the \texttt{im2col} function with 2x2 filter, stride 1, and padding 0.}
\end{figure}

The \texttt{im2col + gemm} implementation is a way to map a convolution as a matrix multiplication (\gls{gemm}), but a data manipulation is needed to accomplish it. The \texttt{im2col} function maps the input matrix into columns to make the Convolution using a \gls{gemm} (Figure \ref{fig:conv2}).

The original Caffe's \texttt{im2col} function is a Penta-loop with dependencies in each iteration.  Therefore, we decided to adapt it to be able to exploit a bit more of parallelism. To create the \gls{phast} version, we merged all the loops and parameterized it with only one index. This change allowed \gls{phast} to use all the available threads as appropriate as each thread is now independent.

\subsubsection{Back-Propagation}
In the feed-forward stage, the im2col function duplicates some values to make a Convolution with a \gls{gemm}. In the back-propagation, we need to apply the reverse step to propagate the gradients to the previous layers. The most important part is the usage of \texttt{col2im} to map the gradients to the size of the input data.

Like in the feed-forward stage, the original implementation is also a Penta-loop, therefore we followed the same approach as before and we merged the loops and parameterized with only one index.

\subsection{InnerProduct} 

\subsubsection{Feed-Forward} 
In neural networks, there is a layer usually known as Perceptron layer (or Dense layer), but sometimes it is also known as InnerProduct, because the output of the layer is the inner product of the inputs with the weights.

\begin{figure}[ht!]
\centering
\begin{minipage}{0.45\textwidth}
\begin{lstlisting}[language=C++, label={lst:inner1}, caption={Caffe version: InnerProduct (CPU only)}]
template <typename Dtype>
void InnerProductLayer<Dtype>::Forward_cpu(const vector<Blob<Dtype>*>& bottom,
    const vector<Blob<Dtype>*>& top) {

  const Dtype* bottom_data = bottom[0]->cpu_data();
  Dtype* top_data = top[0]->mutable_cpu_data();
  const Dtype* weight = this->blobs_[0]->cpu_data();
  
  caffe_cpu_gemm<Dtype>(CblasNoTrans, transpose_ ? CblasNoTrans : CblasTrans,
      M_, N_, K_, (Dtype)1.,
      bottom_data, weight, (Dtype)0., top_data);
  
  if (bias_term_) {
    caffe_cpu_gemm<Dtype>(CblasNoTrans, CblasNoTrans, M_, N_, 1, (Dtype)1.,
        bias_multiplier_.cpu_data(),
        this->blobs_[1]->cpu_data(), (Dtype)1., top_data);
  }
}
\end{lstlisting}
\end{minipage}
\hspace{2em}
\begin{minipage}{0.45\textwidth}
\begin{lstlisting}[language=C++, label={lst:inner2}, caption={PHAST version: InnerProduct (same code for CPU and GPU)}]
template <typename T, unsigned int policy = phast::get_default_policy()>
struct matrixPlusVectorRows : phast::functor::func_vec<T, policy> {
  _PHAST_METHOD matrixPlusVectorRows() {}
  
  _PHAST_METHOD void operator()(phast::functor::vector<T>& row) {
    for(auto r = row.begin(), i = vec.begin(); r != row.end(); ++r, ++i)
      *r += *i;
  }
  phast::functor::vector<T> vec;
};

template <>
void InnerProductLayer<float>::Forward_cpu(const vector<Blob<float>*>& bottom,
  const vector<Blob<float>*>& top) {
  phast::matrix<float> matA = bottom[0]->getDataAsMatrix(M_, K_, false);
  phast::matrix<float> matB = this->blobs_[0]->getDataAsMatrix(K_, N_, !transpose_);
  phast::matrix<float> matC = top[0]->getDataAsMatrix(M_, N_, false);

  phast::dot_product(matA, matB, matC);

  if (bias_term_) {
    matrixPlusVectorRows<float> matrixPlusVectorRows;
    matrixPlusVectorRows.vec.link(this->blobs_[1]->getDataAsVector(N_));
    phast::for_each(matC.begin_i(), matC.end_i(), matrixPlusVectorRows);
  }
  if(!transpose_) matB.transpose();
}
\end{lstlisting}
\end{minipage}
\end{figure}

In this block (see Listing~\ref{lst:inner1}), we find a \gls{gemm}, but going deeper it is noticeable that there is a vector addition over the rows of the matrix. Hence, we implement that functionality in the \texttt{matrix\-Plus\-Vector\-Rows} functor (Listing~\ref{lst:inner2}). Using a \gls{gemm} to represent other operations is that can be found very often in Caffe: its creators have mapped all possible operations to matrix multiplications to make them easier to port to GPU, but for now, we developed specific functors for these operations.
Also, it is noticeable that the code in Listing~\ref{lst:inner1} is CPU-only code, and thus its GPU counterpart is in another source file. Specifically, the original CPU version of Caffe needs other 10 lines of source-code for caffe\_cpu\_gemm() function (total 28), while the original GPU version of Listing~\ref{lst:inner1} is 23 source-code lines plus 27 in the called functions (total 50). Conversely, the code in Listing~\ref{lst:inner2} can be run on both CPU and GPU and is only 27 lines of source-code.

\subsubsection{Back-Propagation}
The InnerProduct back-propagation is a bit different from the other block we have seen. Instead of reversing the operation made in feed-forward to map to each value its own gradient and modify its weight, we added to the weights a scaled gradient based on the original data. Then, we did the same with the bias, and lastly, we propagated the changes to the previous layer. Despite the trick to map all operations to matrix multiplications, this layer is very straightforward.

\subsection{Pooling} 

\subsubsection{Feed-Forward}
The Pooling layer applies a mathematical function to a set of numbers to get only one value, such as maximum, minimum, or mean. Like the Convolution layer, it works using a sliding window over the input data and applying the function to each set.

The structure is very similar to the Convolution block, but this time, we did not apply the same technique. We had only parallelized the outer loop.

\subsubsection{Back-Propagation}
During the feed-forward stage, we stored the origin of each output value. Therefore, we have to map the values from the output to the input using that list of mapped values. Its objective is to apply all the gradients to its corresponding positions.

Like in the feed-forward stage, we did not merge all the loops, because we did not verify that it would have continued to work properly. Therefore, we only parallelized the first loop.

\section{Evaluation}
\subsection{Testbench}
In this work we have used the \gls{phast} library 1.1.1 compiled with GCC 8.3.0 and CUDA 10.1. The Caffe framework was obtained from the official git repository, using the \texttt{99bd\-99795dcdf\-0b1d3086a\-8d67ab1782a\-8a08383} commit. The hardware platform is a high-performance workstation mounting an Intel i9 9900K @ 3.60GHz, 8 cores hyper-threaded and 32 GiB RAM memory, and a Geforce RTX 2080 8GB GDDR6X with NVIDIA driver 430.50. This is running Ubuntu 18.04 with Linux 5.0.0-36-generic.

\subsection{Results}
Firstly, we trained two LeNet-based neural networks using the original Caffe code. The first one, built from 6 layers (2 Convolutions, 2 Poolings, and 2 InnerProducts) was used to classify the MNIST~\cite{mnist} database. The second one was used to classify the CIFAR10~\cite{cifar} database, and it is composed of 8 layers (3 Convolutions, 3 Poolings, and 2 InnerProducts). Additionally, both networks had a SoftMax layer with loss, an Accuracy layer, and at least 1 layer with the ReLU function.

Then, we tested the same examples with our Caffe version ported to \gls{phast}. We run successfully both neural networks in CPU and GPU using \gls{phast}. To check the results, we used the set of inputs that Caffe provides checking several parameters such as the output of the network, the accuracy, the loss, and some intermediate matrices. We preferred to use the intermediate matrices and the outputs to be sure that both versions (the original and ours) were obtaining the same results, despite only with the accuracy and the loss were enough to validate the results. 

Since the same results were found that with the original CPU Caffe's implementation, we claim that all blocks were working properly. Then, we corroborate that our version of the Caffe framework with \gls{phast}, which only has a single code, is able to run on CPU or GPU depending on the Makefile used.

Finally, we decided to test how accurate was our implementation with respect to the original one. Using the Caffe test files, we checked all the blocks ported to \gls{phast} that had a test (Table \ref{table:res}). We noticed that all the functionality ported was working, and only tests that had unimplemented functionality failed.

\vspace{.5em}

\begin{table}[ht]
  \centering
  \begin{tabular}{l|cc|cc}
    Block & Passed & Not Passed & Total & \%Passed\\
    \hline
    Convolution & 3 & 12 & 15 & 20\\
    Pooling & 11 & 0 & 11 & 100\\
    InnerProduct & 9 & 0 & 9 & 100\\
    SoftMax & 4 & 0 & 4 & 100\\
    SoftMax Loss & 4 & 0 & 4 & 100\\
    Accuracy & 9 & 3 & 12 & 75\\
  \end{tabular}
\vspace*{.2cm}
  \caption[Caffe tests results]{\label{table:res} Caffe tests results for the modified blocks in single precision floating point numbers.}
\end{table}

\subsection{Performance}
By now, we have focused on the porting process and checking the functionality of our version. Additionally, some parts are still missing so our current tests are quite preliminary and based on not very big networks. By the way, dealing with small networks implies that the associated data structures managed in the layers to be distributed to, or gathered from, the parallel devices (e.g., GPU) are relatively small and thus more subject to fixed transfer overhead and less able to express the maximum rated steady-state transfer rate. Furthermore, relatively small kernel executions (e.g., some steps elaborate 32 x 16 matrices) intrinsically do not benefit a lot from being distributed onto architectures exposing high-parallelism while, conversely, can suffer the overhead for managing the parallel execution itself. So, overall, a small network can be even considered a non-trivial test case for a framework that aims to exploit automatic parallelization.

Here, Table~\ref{table:res} shows some initial performance results and highlights that the original Caffe CPU code, employing OpenBLAS, outperforms PHAST partially ported version by around 2.8 times, and the GPU original version is around 4.0 times faster than the PHAST one. This is a snapshot of the performance differences at this stage of the porting phase and, in the following, we will discuss the factors that surely contribute to this figures as to highlight the path that we will follow to complete the porting, along with the qualitative expected results. 

\begin{table}[ht]
  \centering
  \begin{subtable}{0.45\textwidth}
  \centering
  \begin{tabular}{l|cc}
     & CPU & GPU \\
    \hline
    Caffe & 71.42 & 7.24 \\
    Caffe (PHAST)\hspace{1cm} & \hspace{0.25cm}198.60\hspace{0.25cm} & \hspace{0.25cm}21.81\hspace{0.25cm} \\
  \end{tabular}
\vspace*{.2cm}
  \caption{MNIST}
  \label{table:res:times1}
  \end{subtable}
  \begin{subtable}{0.45\textwidth}
  \centering
  \begin{tabular}{l|cc}
     & CPU & GPU \\
    \hline
    Caffe & 399.50 & 16.65 \\
    Caffe (PHAST)\hspace{1cm} & \hspace{0.25cm}1113.71\hspace{0.25cm} & \hspace{0.25cm}67.40\hspace{0.25cm} \\
  \end{tabular}
\vspace*{.2cm}
  \caption{CIFAR-10}
  \label{table:res:times2}
  \end{subtable}
  \caption{Average Forward-Backward execution time (ms)}
  \label{table:res}
\end{table}

Among the main direct reasons for this performance difference we expect a major role to be played by the fact that not all layers of the network are ported into PHAST Library yet. However, the \textit{heaviest} layers, like the convolutional ones, have been already ported so we can expect that fixing this point will increase performance of the PHAST version but will not erode the majority of the current difference.

Furthermore, the convolutional algorithm required to be written in PHAST at the user-level because PHAST does not currently provide a native convolutional iteration. Therefore, in the current version of the porting, a quite efficient algorithm was used but, pragmatically, we postponed to a later stage the investigation and implementation of a highly-optimized, state-of-the-art convolutional scan on the data structures. When all the network will be ported we will address this issue and either optimize the algorithm/code or implement it within PHAST Library in a more intimate way with its internals in order to potentially extract some more performance. From our pre-analyses, we expect that the intrinsic acceleration of the convolutional phase will not be huge but potentially meaningful. And, given that this phase accounts for a large fraction of the overall computational execution, it is reasonable to expect that its speedup is translated into a similar overall faster execution.

From another perspective, current porting status implies that a complete porting is highly likely to cause significant speedups due to indirect effects. In fact, having some network layers and framework crucial parts not yet ported (e.g., in ReLu layer the activation function can be expressed by means of PHAST algorithms) induces frequent data transfers, especially in the GPU, between PHAST parallel execution phases and the orchestrating CPU code, as well as, the parallel execution phases performed in the original version of Caffe. So, every time a layer already ported in PHAST is followed by a layer still in the original version, or viceversa, such data transfers need to be done as well as synchronizations between devices and/or phases need to be performed and, surely, this is going to hurt performance. Dealing with relatively small networks as in our test cases, at a certain extent emphasises this situation compared to bigger networks.

We expect the fix to this point to have a major effect in reducing the current performance gap between the original code and the PHAST one. The effect will be indirect and induced by the porting of all layers so that the inference/back-propagation activities will mainly run without artificial interruption across the layers and unneeded data transfers. Surely this will be more evident in the GPU-targeted code because such overhead can be expected to count more. As a rough estimate, we can spot around 10 and 30 unnecessary transfers, for MNIST and CIFAR respectively, between the original and PHAST "domains" in the inference phase only. A similar number, at least, is present in the back-propagation phase. Therefore, this activities are not negligible performance-wise.

To a closer inspection of the overhead induced by the data transfer between ported and not-yet-ported layers, there is another factor that could even constitute the major source of the performance difference in both CPU and GPU target code. In fact, the original Caffe version of the layers relies on OpenBLAS-friendly data structures, i.e., having column-major-order memory allocation, while PHAST is currently written assuming row-major-order in multi-dimensional containers. Therefore, every time the code crosses the boundary between the original version and the PHAST ported one, the involved data structures do need also a conversion from one format to the other, on top of the mere transfer. So, not only a number of data transfers to/from the parallel PHAST execution are unnecessary in themselves, but they require also an additional copy host-side per transfer as to transpose the memory layout.  
Our expectation is that this indirect factor could be the one representing the biggest quote in the current gap breakdown.

This situation, at the current stage of the porting, was unavoidable because it is a direct consequence of following an incremental porting strategy, which, in turn, was required for the controllability and effectiveness of the porting work itself. In fact, this approach allowed us to port one piece at a time of the network, focusing on debugging and performing regression tests iteratively and in a tight controlled fashion as not to risk to have the code diverge. In all the porting phases we have always been able to run and test Caffe execution comparing its output to the original version.

For sure, we expect that once we have ported the entire set of layers, some additional refactoring steps could be performed as to make (some of) the code and (at least some of) the data structures evolve towards a shape more naturally fitting PHAST and, probably, some features that can enable a more effective parallelism exploitation.

Furthermore, some data-structures and parallel activities are quite small and could benefit from merging them into fewer, more complex, kernels to be executed in parallel. However, this benefit could even be quite small, especially for the GPU target, because once all the parallel activities are run on the parallel architecture, the runtime scheduler is potentially capable of very advanced strategies for making the most out of the available hardware resources.

Concluding, the current version of the PHAST partially ported Caffe code, despite being single-source and allowing high coding-productivity and automatic targeting to parallel CPUs and NVIDIA GPUs, exhibits a non-negligible performance slowdown compared to the original versions for the respective architectures. However, we have described how most of the performance gap is likely to be due to the overhead induced by the unnecessary data transfers between the already ported layers and the original ones, both as the intrinsic not-needed transfer time and the time to adapt the memory layout of the transferred data.

As ongoing and near-future work, we aim at completing the porting as to evaluate the exact performance difference between the original and PHAST version of this and other neural networks as well as to start identifying the best parallelization/performance tuning parameters in each kernel. This will be particularly interesting considering that PHAST Library allows exploring such parallelization/tuning space parameters without altering the source code and its correctness as to adapt the application to different instances of the architectures, like different CPUs and especially GPU families and versions. 

\subsection{Lessons Learned}
The first lesson we learned is that current and forthcoming frameworks base their APIs on C++. Thus, a deep understanding of the language, including its newest components (functors, templates, etc) is a key fact that can enable efficiency, robustness and, potentially, portability.

Another important thing is that the porting process is very influenced by the software functionality to be ported. A strong requirement is a high understanding of the original programming model, its purpose, and the specific application domain in which it is applied. It may result in a heavy and complex task for \textit{impedance matching} between application algorithmic structure and the parallel nature of the architecture through the PHAST Library and programming approach. 

Finally, and regarding portability, we have seen very appealing the idea of \emph{a single source-code for many devices}. Virtually any application or framework can be ported. A second step (which was out of the scope of this work) is, however, to complete and polish the porting as to improve overall performance and verify performance portability across architectures.

In this sense, this experience has already proved very valuable from the PHAST standpoint. In fact, some overhead that was initially emerging in PHAST Library applied in the considered application domain, has induced specific improvements in some PHAST base mechanisms for data exchange between the orchestrating CPU and parallel execution devices (e.g., GPU). Specifically, we were able to release a few synchronizations in the inner layers of PHAST Library, which were originally present and that, actually, we now proved to be unnecessary in the specific context of this work and in similar algorithmic situations.

\section{Conclusions}
In this work, we have presented our developments and decisions to port the Caffe library to be used with the PHAST Library. This way, we have shown that it is possible to move from two code sources (one for CPUs and another one for GPUs) to only one single code. This single code is portable and can run on different devices by means of compiling options.

As deep learning and neural networks are very important today, choosing Caffe as a use case seems to be a good choice.

We also have shown that \gls{phast} is suitable to be used in real applications, it is easy to use and remembers well-established highly-expressive techniques. 

There is several work left behind for future developments, among them: \emph{Enhancing the performance}: We have detected some points of improvement in our development, as reducing the number of copies made at each operation or maintaining some structures as long as possible. We plan to implement them in the near future. \emph{Complete the port of Caffe to \gls{phast}}: Apart from reducing the performance gap between the original Caffe implementation and our porting, completing this task will give us the opportunity to release Caffe freely so it can be used by other researchers. \emph{Extend our work to other devices}: Increasing the portability is one of our main goals. By choosing PHAST Library, our version will evolve as soon as new architectures are supported (such as TPUs, for example). This will also give us valuable information for the developing of these programming models.






\section*{Acknowledgements}
This work has been partially funded by the AEI (State Research Agency, Spain) and the ERDF (European Regional Development Fund, EU) under the Contract RTI2018-098156-B-C53. We would also like to thank our laboratory workmates, Francisco Muñoz-Martínez and David Corbalán Navarro, for their fruitful discussions.


\bibliographystyle{splncs04}
\bibliography{ref.bib}

\end{document}